\documentclass[11pt]{article}

\usepackage{colacl}
\usepackage{graphicx}
\usepackage{supertabular}
\usepackage{times}
\def\lra{\leftrightarrow}

\def\hb{\hfill\break}
\title{Using Curvature and Markov Clustering in Graphs for Lexical
Acquisition and Word Sense Discrimination}

\author{
\begin{tabular}{p{3in}p{3in}}
\begin{tabular}{c}
{\small \bf Beate Dorow} \\
{\small \rm Institute for NLP} \\ 
{\small \rm University of Stuttgart, Germany} \\
{\small \rm {\texttt {beate.dorow@ims.uni-stuttgart.de}}} \\
\end{tabular}
&
\begin{tabular}{c}
{\small \bf Dominic Widdows, Katarina Ling} \\
{\small \rm CSLI, Stanford University}, {\small \rm California} \\
\texttt{{\small \rm {\texttt {dwiddows@csli.stanford.edu}}}} \\ \rm\texttt{{\small katarinaling@stanford.edu}}
\end{tabular}
\end{tabular}
\\
\\
\begin{tabular}{p{3in}p{3in}}
\begin{tabular}{c}
{\small \bf Jean-Pierre Eckmann, Danilo Sergi} \\
{\small D\'epartement de Physique Th\'eorique} \\ {\small Universit\'e de Gen\`eve, Switzerland} \\
\texttt{{\small Jean-Pierre.Eckmann@physics.unige.ch}} \\
\texttt{{\small Danilo.Sergi@physics.unige.ch}}\\
\end{tabular}
&
\begin{tabular}{c}
{\small \bf Elisha Moses} \\
{\small Department of Physics of Complex Systems} \\ {\small Weizmann Institute of Science} \\ {\small Rehovot, Israel} \\
\texttt{\small fnmoses@wicc.weizmann.ac.il} \\
\end{tabular}
\\
\\
\end{tabular}
}

\begin{document}
\maketitle

\begin{abstract}
We introduce two different approaches for clustering semantically 
similar words. We accommodate ambiguity by allowing a word to belong to several clusters.

Both methods use a graph-theoretic representation of words and
their paradigmatic
relationships.  The first approach is based on the
concept of {\em curvature} and divides the word graph into classes of
similar words by removing words of low curvature which connect several
dispersed clusters.

The second method, instead of clustering the nodes,
clusters the links in our graph. These contain more specific contextual
information than nodes representing just words. In so doing, we
naturally accommodate ambiguity by allowing multiple class membership.

Both methods are evaluated on a lexical acquisition task, using
clustering to add nouns to the WordNet taxonomy. The most effective
method is link clustering.
\end{abstract}

\section{Introduction}

Graphs have been widely used to model many practical situations
\cite{chartrand-graph}, including many semantic issues: 
The link structure of the World
Wide Wed has been investigated and manipulated to detect shared
interest communities \cite{Eck02}, and modeling WordNet as a graph
has yielded insight about semantic relatedness and ambiguity
\cite{Sig02}.

In this paper, we present a graph model for nouns and paradigmatic
relationships collected from the British National
Corpus (BNC)\footnote{\texttt{http://www.natcorp.ox.ac.uk/}} using simple
lexicosyntactic patterns. 

The resulting semantic structure can be used for lexical acquisition
(by gathering nodes into clusters and labeling the clusters) and word
sense discrimination (by determining when a node in the graph is really a
conglomeration of nodes representing different senses). 

We introduce two tools to approach these tasks: the curvature measure
of \cite{Eck02} and the Markov Clustering (MCL) of \cite{Don00}.  The
first algorithm removes the nodes of low curvature (the hubs of the
graph), upon which the word graph breaks up into disconnected
coherent semantic clusters. MCL decomposes the word
graph into small coherent pieces via simulation of random walks in
the graph which eventually get trapped in dense regions, 
the resulting clusters.

Both methods effectively place each node into exactly one
cluster, breaking the graph into equivalence classes. The
shortcomings of any such approach become apparent once we consider
ambiguity---when each word is treated as an indivisible unit in the
graph, we need to split these semantic atoms to account for different
senses.  We investigate an alternative approach which treats each
individual coordination pattern as semantic node, and agglomerates
these more contextual units into usage clusters corresponding closely
to word senses.

A comparative evaluation of these methods on a lexical acquisition
task is presented in Sect.~\ref{evaluation-sec}.

\section{\label{sss}The graph model}

\begin{figure}
\begin{center}
\includegraphics[width=6.6cm]{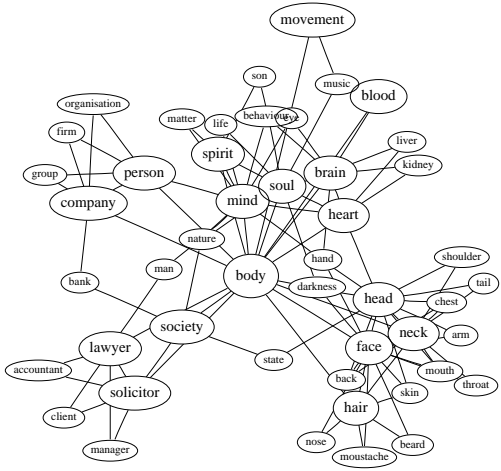}
\caption[body]{\label{body} Local graph around {\em body}}
\end{center}
\end{figure}

To build a graph representing the relationships between nouns, we used
simple regular expressions to search the BNC, which is tagged for parts of 
speech, for examples of lexicosyntactic patterns which are often indicative 
of a semantic relationship \cite{hearst-hypernyms}. 
The hypothesis is that nouns in coordinations are semantically similar
(cf.~\newcite{Ril97}, \newcite{Roa98}, \newcite{Wid02}). 
We collected coordinations of noun phrases using simple patterns, dropped 
prenominal modifiers, and built a word graph by \hfill\break
~~1.~Introducing a node for each of the nouns;\hfill\break
~~2.~Connecting two nouns by an edge if they co-occurred 
in a coordination. 
\hfill\break
Consider the following example sentences drawn from the BNC
containing a coordination ``{\em body}'' appearing in: 
\begin{quote}
{\small Legend has it that the mandarin was so grateful to Earl Grey for 
services rendered that he gave him his secret tea recipe, to keep {\em 
mind, body and spirit} together in perfect harmony. 

So far as ordinary citizens and non-governmental bodies are 
concerned, the background principle of English law is that a {\em 
person or body} may do anything which the law does not prohibit. 

Christopher was also bitten on the {\em head, neck and body} 
before his pet collie Waldo dashed to the rescue. }
\end{quote}
The highlighted coordinations give rise to edges 
\begin{quote}
\begin{small}
{\em body$\lra$mind}, {\em body$\lra$spirit}, {\em mind$\lra$spirit} 

{\em body$\lra$person} 

{\em body$\lra$head}, {\em body$\lra$neck}, {\em head$\lra$neck} 
\end{small}
\end{quote}
in the word graph. Fig.~\ref{body} displays a 
subgraph of our word graph centered around {\em body} and consisting of 
the top $17$ neighbors of {\em body} and the top $8$ neighbors of the 
neighbors. 
The word graph has this very simple interpretation: Words which are directly 
linked are semantically close. 
%%%If, on the other hand, many links have 
%%%to be traversed to go from one word to another, then those words are 
%%%less similar than another pair of words that are fewer links away from 
%%%each other. 
%%%
This graph consists of $88,900$ nodes (word types) 
and $551,745$ edges. 
We ignore the order in which two words co-occur in a coordination, the edges 
in our graph are not given any direction. 
%%%%\footnote{We are aware of the problem posed 
%%%%by idiomatic expressions such as ``Elephant and Castle'' and hypernymic expressions 
%%%%such as ``schools, colleges, hospitals and other institutions'' in which the order of 
%%%%the constituents is (at least partially) fixed, and we plan to deal with this problem 
%%%%in the future.}
%%%%
To reduce noise, we keep only those links in the graph which appear in a triangle, since the links within a triangle confirm each other's significance. This results in a reduced 
word graph consisting of $48,727$ nodes and $505,412$ edges.

\section{Graph curvature and quantifying semantic ambiguity}

Our approach to assessing ambiguity is similar to the one proposed by \newcite{Spr98}, in that our measure also quantifies ambiguity based on the semantic 
cohesiveness of the target word's neighborhood. Words with a very tightly-knit 
neighborhood are assigned smaller ambiguity scores than words whose neighborhood 
is rather fuzzy. 

\begin{figure}
\begin{center}
\def\wdd{2.6cm}
\includegraphics[width=\wdd]{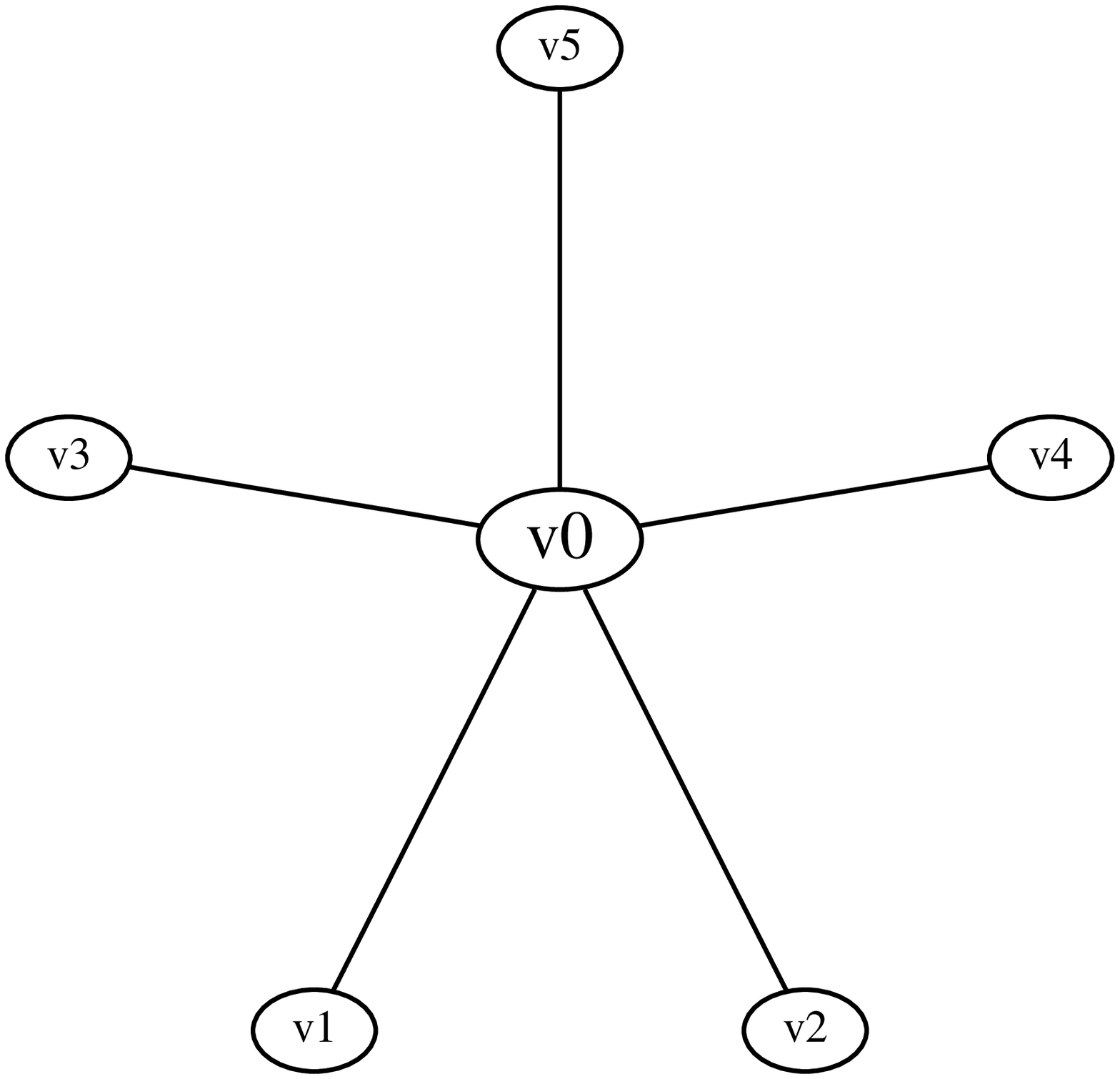}~~~~~~~~~~\includegraphics[width=\wdd]{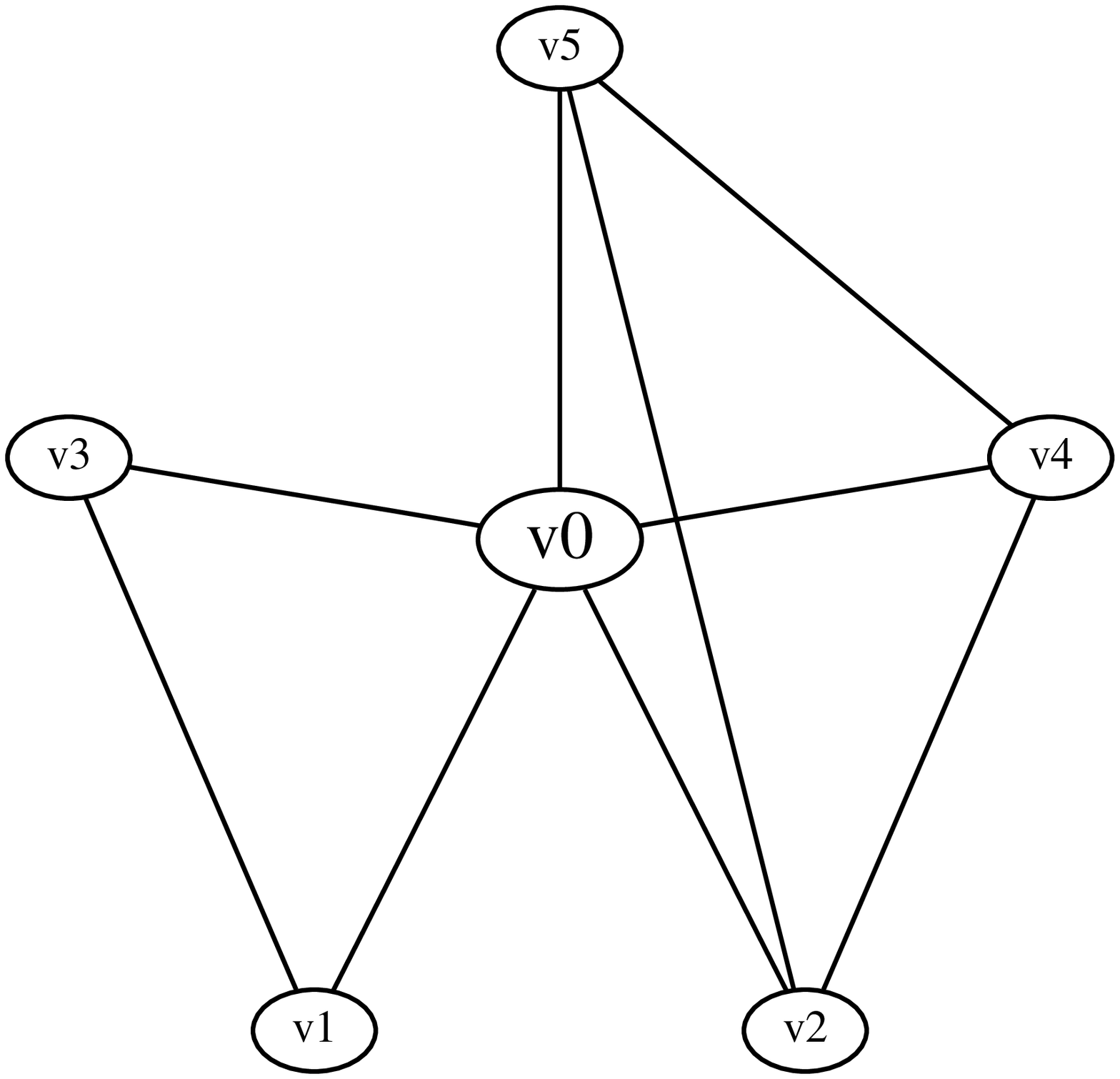}\\
\includegraphics[width=\wdd]{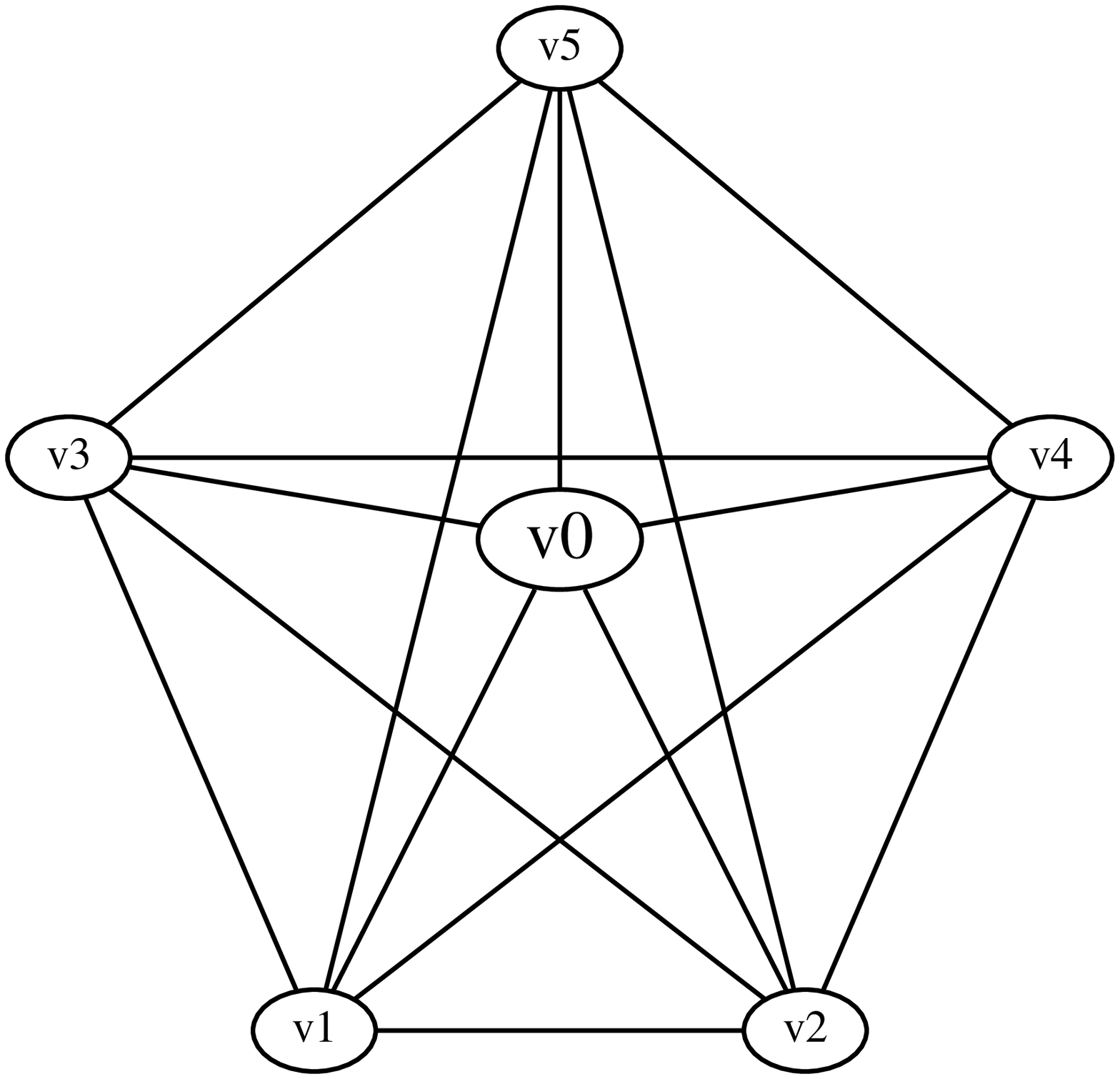}
\end{center}
\caption[curv_0]{\label{curv_0} \small Left: $\mbox{curv}(v_0)=0$, Right:
  $\mbox{curv}(v_0)=0.3$, Bottom:  $\mbox{curv}(v_0)=1$.}
\end{figure}

We measure the semantic cohesiveness of a word's neighborhood (and as a result 
ambiguity) as the {\em curvature} of the word in the graph. Curvature is 
a property of nodes in a graph which quantifies the interconnectedness of a 
node's neighbors. The curvature $\mbox{\small curv}(w)$ of a node $w$
is defined by: 
\begin{eqnarray*}
\mbox{\small curv}({\small w}) = \frac{\mbox{\small \#(triangles $w$ participates
    in)}}{\mbox{\small \#(triangles $w$ could participate in)}} 
\end{eqnarray*}
Curvature is the fraction of existing links among a node's neighbors out 
of all possible links between neighbors. It assumes values between $0$
and $1$. A value of $0$ occurs if there is no link between any of the
node's neighbors (i.e. the neighbors are maximally disconnected), and
a node has a curvature of $1$ if all its neighbors are linked
(i.e. its neighborhood is maximally connected). 
Fig.~\ref{curv_0} shows nodes of low, medium and high curvature respectively.
Curvature measures whether neighbors of a word are neighbors of each other.
Very specific 
unambiguous words have high curvature, because they usually live in small, 
semantically very cohesive communities in which many pairs of nodes have 
mutual neighbors. These communities thus contain a high density of 
triangles. Examples for tight word communities are the days of the
week, the world religions, Greek gods, 
chemical elements, English counties, the planets, 
the members of a rock band, etc. 
Ambiguous words, on the other hand, are linked to members of different 
communities (corresponding to the different meanings of $w$) which do
not know each other. An ambiguous word's neighborhood thus has a low 
density of triangles which results in a low curvature value. 

In information theory, it is common to use the negative logarithm of 
relative word frequency to measure a word's information content ($\mbox{\small info}(w) = -\mbox{\small log}(\mbox{\small rf}(w))$). The intuition is
that very frequent words tend to be very general and uninformative, and that very 
infrequent words tend to be more specific.
Among the most frequent words in our model are countries, which according 
to ${\small \mbox{\small info}(\cdot)}$ would be wrongly categorized as very uninformative, ambiguous
words. 

\begin{figure}
\begin{center}
\includegraphics[width=6.6cm]{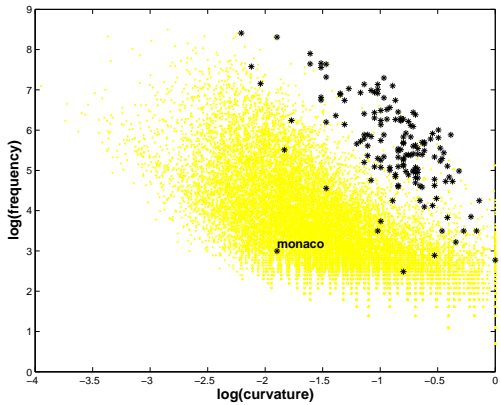}
\caption[countries]{\label{countries} Curvature vs. model frequency}
\end{center}
\end{figure}

\begin{figure}
\begin{center}
\includegraphics[width=6.5cm]{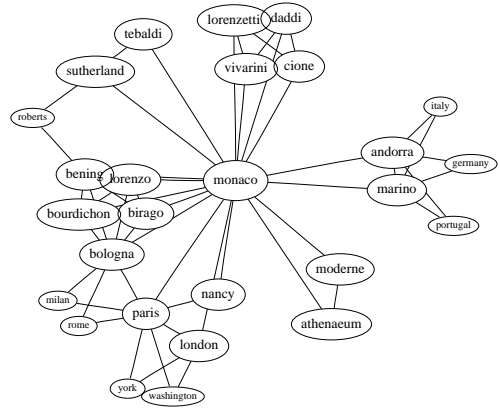}
\caption[monaco]{\label{monaco} Local graph around {\em monaco}}
\end{center}
\end{figure}

Fig.~\ref{countries} is a plot of curvature against frequency in our model. 
The countries among the nodes are indicated by black stars. Very clearly, 
the curvatures of the countries are considerably higher than the average curvature of words
with similar frequency in the model, suggesting that, despite their
high frequency, they are all very informative, i.e. unambiguous. 
The outlier in the lower left corner of the plot is {\em monaco} which 
may not seem ambiguous, but which has several different meanings in the BNC: 
country, city, 14th century painter and 20th century tenor (cf.~Fig.~\ref{monaco}).
%Another country receiving a very low curvature score is 
%{\em georgia} which correctly reflects the high ambiguity between the 
%country and American state (cf.~Fig.~\ref{georgia}).
%\begin{figure}
%\begin{center}
%\includegraphics[width=4.5cm]{./images/georgia}
%\caption[georgie]{\label{georgia} Local graph around {\em georgia}}
%\end{center}
%\end{figure}
To check how well curvature is suited for detecting and assessing ambiguity, 
we take all words in our model which are listed in WordNet and check how 
strongly curvature and the number of WordNet senses are related. Since the 
relationship does not have to be linear, we replace curvature and number of 
WordNet senses by their ranks before computing the Pearson correlation 
coefficient. 
We also wanted to see whether and to which degree curvature better reflects 
ambiguity than a word's frequency in the model or its {\em degree} (the number of
links attached to a node) in the graph.
Table \ref{correlations} lists the mutual Pearson correlations between any two quantities 
out of model frequency, degree, curvature and number of WordNet senses. Our analysis 
shows that with a negative correlation of $-0.538$, curvature is more strongly related to 
the number of WordNet senses and thus a better measure of ambiguity than model
frequency or degree. 
\begin{table}[t]
\begin{scriptsize}
\caption[correlations]{\label{correlations} Rank correlations between any two out of number
  of WordNet senses, word frequency in the model, degree and curvature.}
\begin{center}
\begin{tabular}{c|cccc}
   &  senses & freq & deg & curv\\
   \hline
   senses & 1.000 & 0.475 & 0.480 & -0.538\\
   freq & & 1.000 & 0.963 & -0.865\\
   deg & & & 1.000 & -0.884\\
   curv & & & & 1.000
\end{tabular}
\end{center}
\end{scriptsize}
\end{table}
\section{Inducing classes of similar words}

A semantic category (also referred to as a semantic field) is a grouping of vocabulary 
within a language, organizing words which are interrelated and define each other in various ways.
The acquisition of semantic categories from text has been addressed in
several different ways: Work in this direction can be found in
(\newcite{Per93}, \newcite{Sch98}, \newcite{Pan02}, \newcite{Dor03}).

Word clustering techniques differ in the way they assign words to clusters, either allowing 
words to belong to several clusters (soft clustering), or assigning words to one and only one 
cluster (hard clustering). 
%%\subsection{Hard versus soft clustering} 
%%\label{hard_vs_soft}
%%Clustering is a method for dividing a set of objects into homogeneous 
%%groups such that objects within a group are as similar as possible 
%%and objects of different groups are maximally dissimilar. 
%%%By lumping together objects within clusters, the initially large and 
%%%complex data set is summarized by a small number of latent prototypes 
%%%giving it a simpler, clearer and more robust representation.
%%In hard clustering, each object is assigned to exactly one cluster.
%%Soft clustering algorithms, on the other hand, allow elements to
%%%appear in several clusters at the same time. 
A problem of hard 
clustering techniques is that each word is coerced into a single 
cluster irrespective of whether it is closely associated with other 
clusters, too.

Semantic categories overlap considerably, but hard clustering produces
mutually exclusive clusters and forces ambiguous words to associate with 
a single cluster only. We therefore concentrate on soft clustering.

\subsection{Graph clustering}

In the following, we describe two approaches to soft clustering of words
in our graph.

{\bf{Curvature clustering}}:
In our word graph, ambiguous words function as bridges between
different word communities, e.g. {\em cancer} is the
meeting point of the animal community, the set of lethal diseases and 
the signs of the zodiac. 
By removing these ``semantic hubs'', the graph decomposes into small 
pieces corresponding to cohesive semantic categories.
In detail, the method for extracting clusters of similar words is the
following:\hb 
1. Compute the curvature of each node in the graph.\hb
2. Remove all nodes whose curvature falls below a certain threshold
  ($0.5$).\hb
3. The resulting connected components constitute clusters of
  semantically similar words.

Application of this algorithm to our word graph results in 700
clusters of size $\ge2$. The resulting clustering
covers $2,306$ of the nouns in our model with $21,218$ of the nodes not
making the $0.5$ curvature threshold and $25,203$ isolated nodes.

This method produces a hard clustering of the high
curvature words. Since high curvature words have a well-defined
meaning, we expect a hard clustering approach to detect the (unique)
semantic category each of these words belongs to.

Curvature clustering in this form cannot give
information on the semantically fuzzy low
curvature words. 
Therefore, we augment each of the clusters with the nodes attached
to it.
Table \ref{example_clusters} lists some of the enriched
clusters. The original cluster (the core of the extended cluster) is
printed in bold font, cluster neighbors which did not pass the
curvature threshold are highlighted in italics, and neighbors which
were isolated in the initial clustering are printed in normal font. 
Often, the core words of high curvature are quite specific and
unambiguous, suggesting that high curvature is a desirable property
for `seed words'  (as in \cite{Roa98}) used for this purpose.
By extending the core clusters to their neighbors, coverage could be
increased to $7,532$ nodes in the graph.

\begin{table}
\begin{scriptsize}
\caption[example_clusters]{\label{example_clusters} Clusters resulting from the curvature approach}
\begin{center}
\begin{tabular}{|p{75mm}|}
\hline
{\bf applewood fruitwood} {\em cherry ivory pine oak}\\
\hline
{\bf jainism sikhism vaisnavism} {\em islam buddhism hinduism christianity judaism}\\
\hline
{\bf horseflies lacewings} {\em butterfly mosquito
    beetle centipedes ladybird bird moth}\\
\hline
{\bf freestyle backstroke} {\em butterfly race medley}\\
%\hline
%{\bf adenine uracil guanine thymine}\\
\hline
{\bf printmaker ceramicist} {\em sculptor painter draughtsman artist}\\
\hline
{\bf pomelo papaya} {\em banana potato pineapple mango peach palm pear parsnip}\\
\hline
%{\bf guadeloupe guiana} {\em venezuela martinique america barbados}\\
%\hline
%{\bf hindi bengali urdu farsi gujarati kannada} {\em aborigines english north}\\
%\hline
{\bf poliomyelitis tetanus} {\em tb kinase cough polio diphtheria malaria disease tuberculosis pertussis} anthrax\\
\hline
{\bf thiamin niacin} {\em riboflavin fibre protein iron calcium}\\
\hline
{\bf oratorio cantata} {\em concert baroque opera aria motet play}\\
\hline
{\bf morphine methadone} {\em chloroform heroin caffeine length phosphate cocaine lsd} librium metabolite\\
\hline
%{\bf hypnotherapy autosuggestion} {\em psychotherapy exercise meditation therapy counselling analysis}\\
%\hline
{\bf stepsister stepbrother} {\em friend father sister stepmother brother}\\
\hline
{\bf insectivores artiodactyls ungulates} {\em mammal herbivore individual carnivore rodent horse order fruit}\\
\hline
{\bf cosine tangent} {\em area sine torsion factor}\\
\hline
\end{tabular}
\end{center}
\end{scriptsize}
\end{table}

{\bf{Markov Clustering}}:
A very intuitive graph clustering algorithm is {\em Markov Clustering}
(http://micans.org/mcl/) developed by \newcite{Don00}. Markov
Clustering (MCL) partitions a graph via simulation
of random walks.
The idea is that random walks on a graph are likely to get stuck
within dense subgraphs rather than shuttle between dense subgraphs via
sparse connections.

%%%The MCL algorithm proceeds by iteratively expanding random walks which allows 
%%%them to depart farther and farther from where they started. Each time the 
%%%random walks are expanded, the transition probabilities between nodes are
%%%re-scaled in such a way that likely transitions are further reinforced and less
%%%frequented transitions are further discriminated against.
%%%As random walks stretch out, bonds within dense regions grow
%%%stronger and connections between dense regions die away.
%%%The transition probabilities converge to an equilibrium state which
%%%can be interpreted as a clustering of the graph: positive
%%%transition probabilities give rise to links between nodes and
%%%transition probabilities of $0$ correspond to a missing link. The
%%%resulting connected components form a clustering of the original graph.
%%%
MCL computes a hard clustering. The nodes in the graph are divided
into non-overlapping clusters. Thus, nodes between dense regions
will appear in a single cluster only, although they are attracted by
different communities. 
Inspired by  Sch{\"u}tze's
method \cite{Sch98} 
we next replace clustering of word
{\em strings} by clustering of word {\em contexts}.

\subsection{Clustering the {\em link graph}}
\label{mcllink}

We consider pairs of words which we linked earlier, as word contexts. For example, {\em organ}
occurs in contexts
 ({\em organ}, {\em piano}), ({\em organ}, {\em harpsichord}), 
({\em organ}, {\em tissue}) and ({\em organ}, {\em muscle}). 
In contrast to the semantic ``fuzziness'' of {\em organ}, each of its contexts
has a sharp-cut meaning and refers to exactly one of the senses of
{\em organ}. 
By clustering word contexts as opposed to clustering the words
themselves, a word's different meanings can be distributed across
different clusters which are then interpreted as word senses. E.g. we can assign ({\em organ}, {\em piano}) and ({\em organ},
{\em harpsichord}) to one context cluster, and ({\em organ},
{\em tissue}) and ({\em organ}, {\em muscle}) to another different context cluster.

\begin{figure}
\begin{center}
\includegraphics[width=2cm]{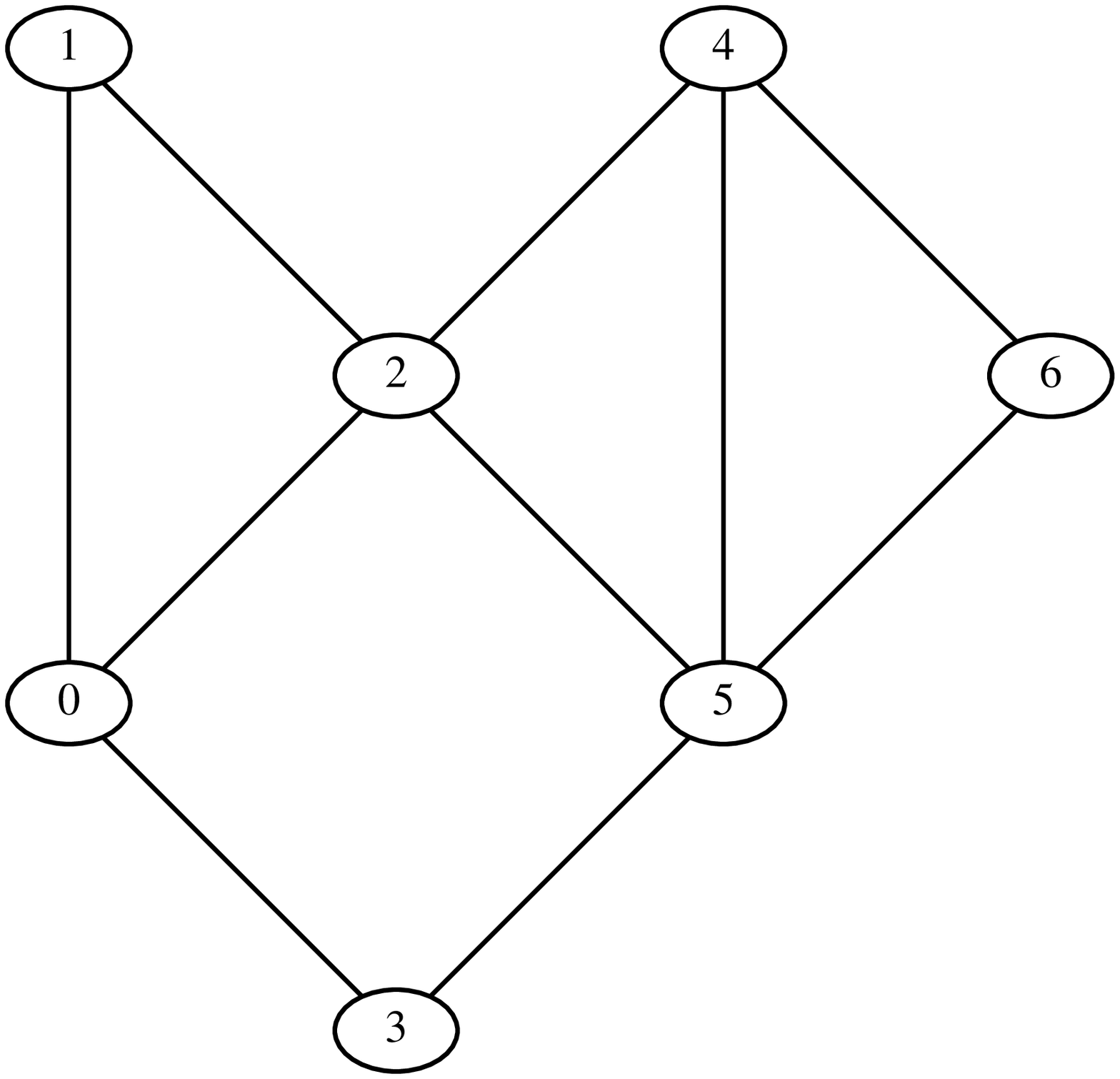}~~
\includegraphics[width=2cm]{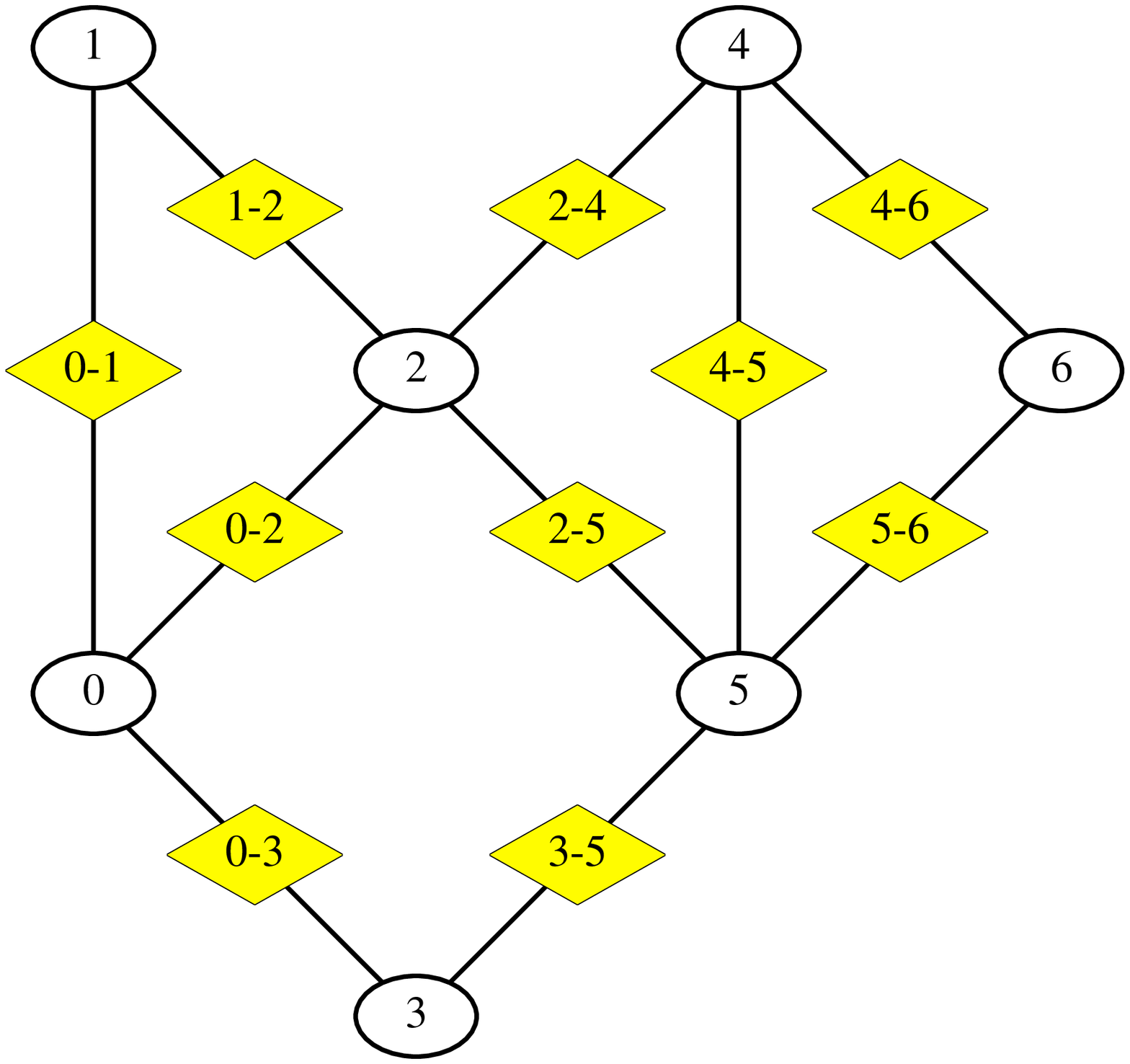}\\
\includegraphics[width=2cm]{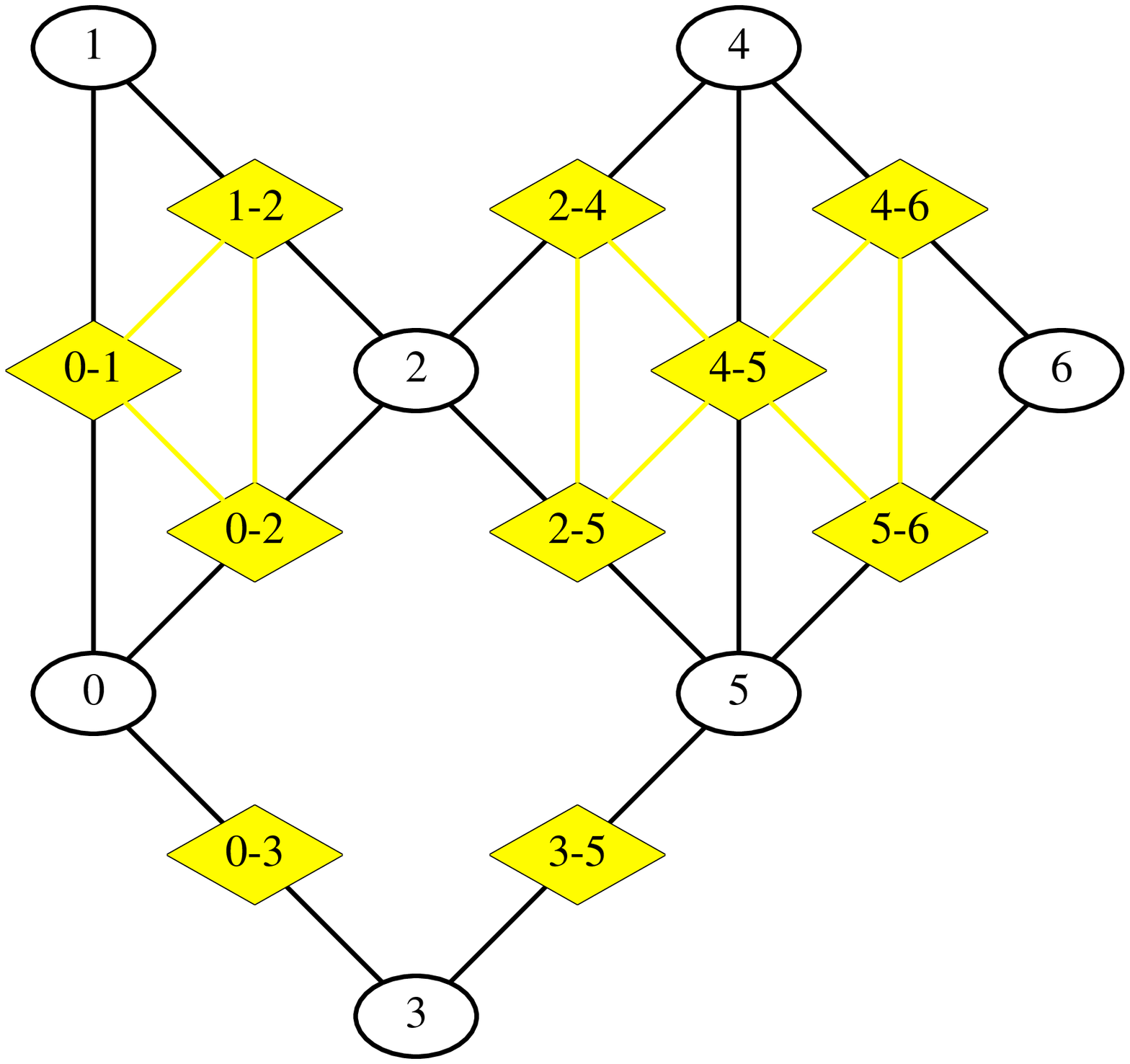}~~
\includegraphics[width=2.4cm]{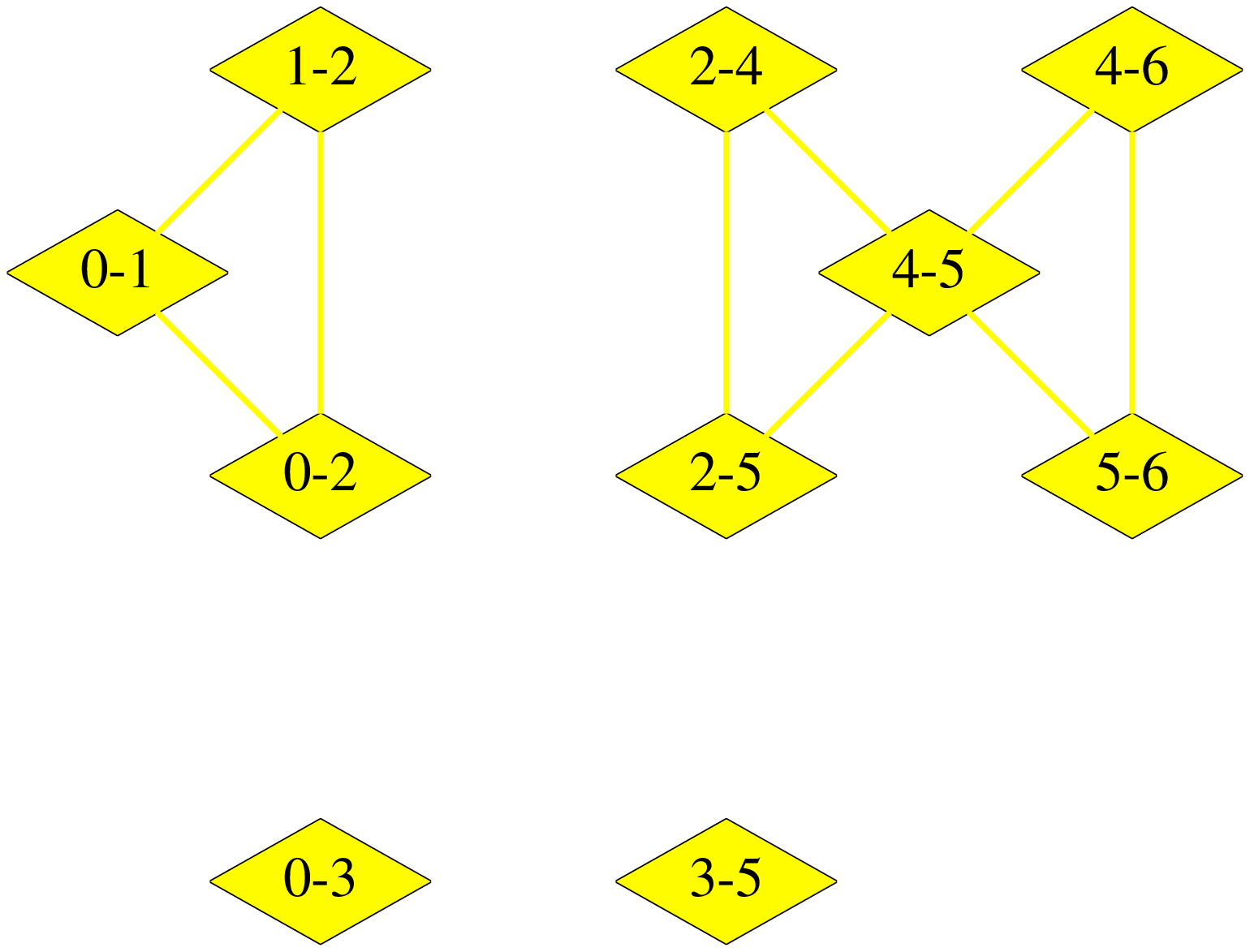}
\caption[link_graph]{\label{link_graph} From $G$ to $G^{\prime}$. The
  original graph, new nodes $n_l$ , new links, the graph $G'$.}
\end{center}
\end{figure}

In the setting of Sect.~\ref{sss}, {\em words} correspond to {\em nodes} in the word graph and
{\em word contexts} coincide with the graph's {\em edges} (with each edge being 
a context of the two nodes it joins). 
We now consider {\em edges} as the fundamental nodes of the {\em link
  graph} $G'$, and define the edges of $G'$ as follows:
We construct the word graph's associated {\em link graph},
$G^{\prime}$, by (see Fig.~~\ref{link_graph}):\hfill\break
~~1. Introducing a node $n_l$ for each link $l$ in the original graph
  $G$.\hfill\break
~~2. Connecting any two nodes $n_{l_1}$ and $n_{l_2}$ in $G^{\prime}$ if $l_1$ and $l_2$ co-occurred in a triangle in $G$.

The two component words $u$ and $v$ of a context $l=(u,v)$ disambiguate each other, e.g. in the ({\em organ}, {\em harpsichord}) context, both {\em organ} and {\em harpsichord} are {\em instruments}, since this is the intersection of all the possible meanings of {\em organ} and all the possible meanings of {\em harpsichord}. The nodes $n_l$ introduced in step 1 therefore have a much narrower meaning than the nodes in $G$.
%\marginpar{\tiny not so easy is regular polysemy; a context ({\em fiat}, {\em ford}) doesn't disambiguate between {\em car} and {\em company}}
 
The links of a triangle in $G$ constitute mutually overlapping word contexts. We therefore expect the links in such a context triangle to have the same ``topic'', and the nodes at the corners of the triangle to have the same meaning. This means, step 2 connects two nodes $n_{l_1}$ and $n_{l_2}$ if the corresponding contexts $l_1$ and $l_2$ are semantically similar.

\begin{figure}[Ht]
\begin{center}
\includegraphics[width=6cm]{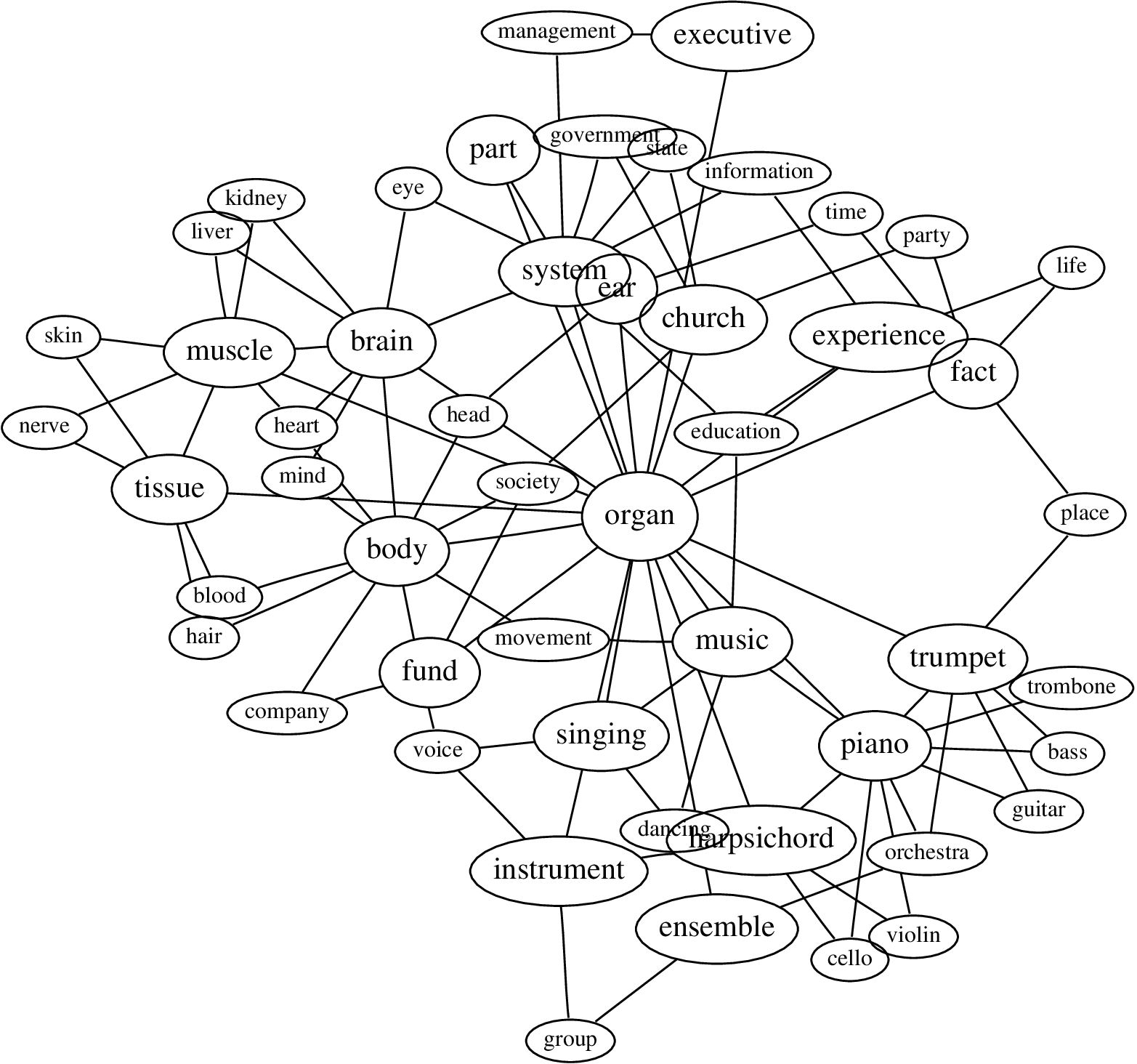}
\caption[organ]{\label{organ} Local word graph around {\em organ}}
\end{center}
\end{figure}

\begin{figure}
\def\wdd{3.92
cm}
\includegraphics[width=\wdd]{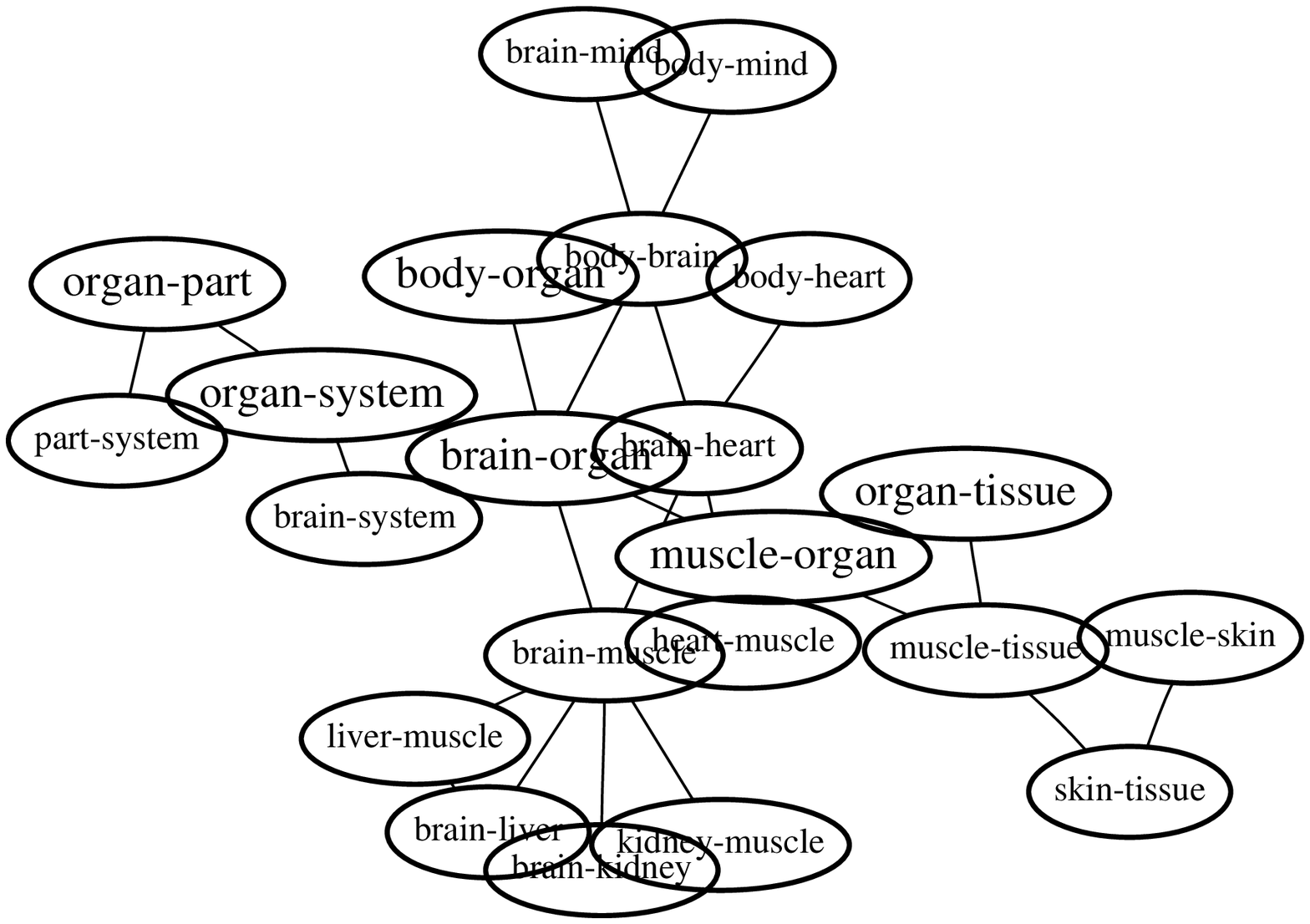}
\includegraphics[width=\wdd]{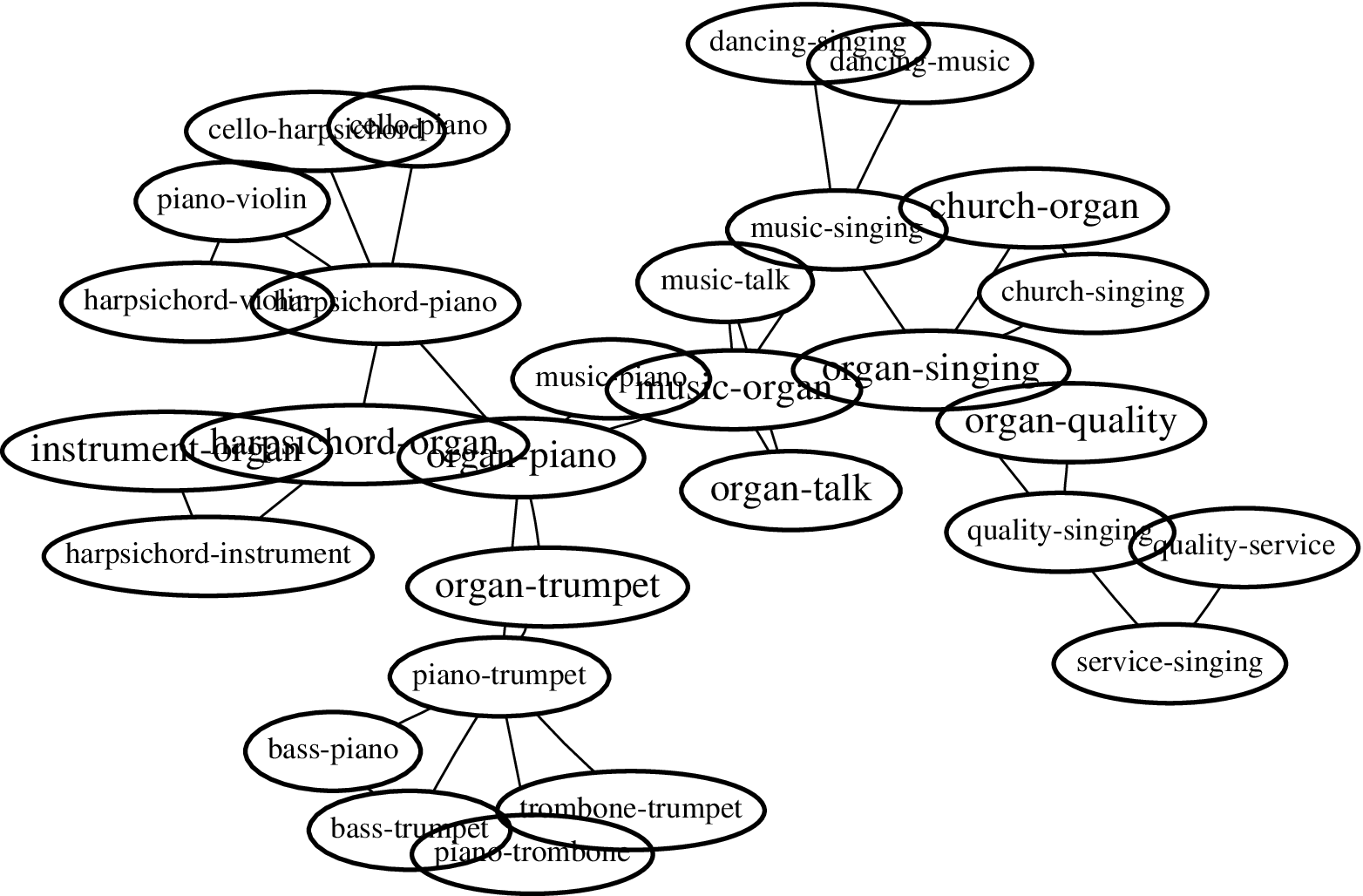}
\caption[organ_dual]{\label{organ_dual} Its associated link graph}
%\end{minipage}
\end{figure}

Fig.~\ref{organ} shows the local word graph around {\em organ}. Its
associated link graph is illustrated in Fig.~\ref{organ_dual} (only
those connected components containing {\em organ} with more than one node are displayed). Note that in the link graph, neighbors corresponding to different senses of {\em organ} are no longer linked.

Instead of clustering words by partitioning the original graph $G$, we cluster word contexts by partitioning $G$'s associated link graph $G^{\prime}$. The nodes $n_l$ in $G^{\prime}$ are built with contextual information, and thus typically have a clear-cut meaning. With little (if any) ambiguity left in the link graph, a hard clustering algorithm, such as MCL, is fit for dividing the contexts into (non-overlapping) similarity classes. 
In detail, our algorithm is:\hb
1) Start with the original graph $G$.\hb
2) Construct the associated link graph $G^{\prime}$.\hb
3) Apply Markov Clustering to $G^{\prime}$.\hb
4) Merge clusters whose overlap in information exceeds a certain threshold.

The clustering resulting from step 3 is
too fine-grained. Several of the context clusters describe the
same ``topic''. We collapse these multiple clusters via another application of MCL, this time applied to a graph of context clusters which are linked if their shared information content (the negative logarithm of the probability of the words they have in common) exceeds $50\%$ of the information contained in the smaller of the two clusters.
Step 4 reduced the $12,786$ clusters resulting from step 3 to a
total of $5,849$ clusters.
%%%%%%%%%%%%%%%%%%%%%%%%%%%%%%%%%%%%%%%%%%%%%%%%%%%%%%%%%%%%%%%%%%%
%%%%%%%%%%%%%%%%%%%%%%%%%%%%%%%%%%%%%%%%%%%%%%%%%%%%%%%%%%%%%%%%%
%%%%%%%%%%%%%%%%%%%%%%%%%%%%%%%%%%%%%%%%%%%%%%%%%%%%%
%\end{center}
%\marginpar{include some example clusters an ambiguous word appears in,
%  e.g. rock, rape, tea, sentence, etc.}
%%%%%%%%%%%

\section{Comparative evaluation for lexical acquisition}
\label{evaluation-sec}

One of the principal uses of word clustering techniques is to supply
missing lexical information. For example, the hypernyms of a word $a$ can
often be inferred from the hypernyms of its neighbors $b_1, \ldots,
b_n$. This property was used by \cite{hearst-customizing} and
\cite{Wid03} to map unknown words into the WordNet taxonomy. The
accuracy of such methods depends on the taxonomy in question, the
method used to obtain the neighbors $b_1, \ldots, b_n$, and the
specificity of the result desired.

From the subset of nouns in our test graph known to WordNet,
we randomly picked a set of 1,200 test words consisting of 600 proper
nouns and 600 common nouns. Each of these two subsets is further
divided into 3 frequency categories (top (500-1000), mid (250-1000),
low (below 250)) which consist of 200 words each. 

Pretending that we don't know the test words, we test how well we
do in re-mapping them into WordNet. 
For each of the test words $t$, we look up which clusters it appears in
and keep its most similar cluster $c_{\max}$. Similarity between words and 
clusters is computed using cosine similarity between their vector representations 
in a vector space model \cite{Dee90}.
%To compute the similarity between a test word $t$ and each of the clusters
%it appears in, we use a variant of the vector space model \cite{Dee90}
%which represents a word as a vector by counting co-occurrences with a set of
%content-bearing words. A cluster (or more generally a set of words)
%can be represented in the same vector space by simply adding up the
%co-occurrence vectors of its constituents. We compute semantic
%similarity between a test word and a cluster it belongs to, by
%comparing the corresponding vectors by means of cosine similarity.

We then assign a sense label to $c_{\max}$ using the sense-labeling
algorithm proposed in \cite{Wid03} which treats any hypernym of any of
the cluster members as a potential cluster label. Potential cluster labels are rated based on 
two competing criteria:\hb
%\begin{itemize}
$\bullet$~The more cluster members a label subsumes, the
better (favoring more general labels).\hb
$\bullet$~The more informative the label, the better (favoring more specific labels).\hb
%The closer (in the WordNet hierarchy) the label to the cluster members the better (favoring more specific labels).
%\end{itemize}
Since including the test word $t$ in the sense-labeling process would be using information about $t$ which we are not given in a real lexical acquisition situation, we disregard both, the test word and the cluster members
which are morphologically related to the test word.

The labeling algorithm outputs a cluster's top five
labels together with a score assessing their adequacy.
We compare each of these labels with each of the test word's ancestors
in WordNet, and, in case of a match, record the number of intervening
levels between the test word and the label. 
E.g. the test word {\em opera} appears in the
cluster ({\em jazz, music, festival, sound, beat, reggae, soul, ballet, funk, country, orchestra, film, table, poetry}).
Table \ref{cluster_labels} shows the cluster labels and scores assigned by the
class-labeling algorithm.
Column {\em match} lists the number of intervening WordNet levels
between {\em opera} and each of the labels.

If a test word $t$ is not covered by the clustering, we do a depth-first search on the original word graph starting at $t$ and moving along the strongest link until we reach a node $t^{\prime}$ covered by the clustering. We then pretend that $t$ belongs to the cluster(s) $t^{\prime}$ appears in.
\begin{table}[htpb]
\begin{scriptsize}
\begin{center}
\caption[cluster_labels]{\label{cluster_labels} Labels assigned to the
  {\em opera} cluster}
\begin{tabular}{|l|c|c|}
\hline
  {\small label} & {\small score} & {\small match} \\
  \hline
  \hline
  {\small auditory communication} & {\small 0.438} & {\small 4} \\
  \hline
  {\small communication} & {\small 0.076} & {\small 5} \\
  \hline
  {\small abstraction} & {\small -0.265} & {\small 8} \\
  \hline
  {\small relation} & {\small -0.500} & {\small 7} \\
  \hline
  {\small social relation} & {\small -0.603} & {\small 6} \\
  \hline
\end{tabular}
\end{center}
\end{scriptsize}
\end{table}
To summarize, evaluation consists of the following steps.
For each test word $t$,\hb
~1.~If $t$ doesn't appear in any cluster, follow strongest links until
you reach a word $t^{\prime}$ which is covered by the clustering and
substitute $t$ with $t^{\prime}$.\hb
~2.~Collect the clusters $t$ appears in.\hb
~3.~Compute the similarity between $t$ and each of the clusters and keep
  only the cluster $c_{\max}$ which is most similar to $t$.\hb
~4.~Compute a class label $l$ for $c_{\max} \setminus \{t\}$.\hb
~5.~Check if (and how closely) $l$ corresponds to one of $t$'s WordNet senses.\hb

\paragraph{Basis for comparison}We use the following simple sense-labeling method as basis for comparison.
For each test word $t$, we find its
nearest neighbor $n$ in the graph. For all the hypernyms of $t$ and $n$, we
find their common subsumer $\mbox{cs}(t,n)$ which minimizes the average distance to $t$
and $n$. 
We are directly using taxonomic knowledge about our test word $t$ to find the 
optimal position in the WordNet tree where $t$ and $n$ should join.
In a real lexical acquisition situation, of course, such information is not available.
This method therefore forms a simplest upper bound on how well we could expect to do in 
mapping unknown words into the WordNet taxonomy.

\paragraph{Results} Table \ref{table_of_results_nn1} 
summarizes the performance of the algorithms 
on the lexical acquisition task described above (similar results are
obtained for proper nouns).
For each test set and each method, row N lists the
number (percentage) of test words which are not in WordNet. The number
(percentage) of words which received a label not corresponding to any
of its senses with any number of intervening WordNet levels is listed
in row W. The rows $i=1..12$ contain the number of words which were
assigned a correct label with $i$ or less intervening WordNet
levels. For these rows, the percentages in parentheses are relative
to the total number of words which were assigned a correct label.

Naturally, the baseline method, which is using (normally unknown) taxonomic information about the
test word itself, performs best. Of all the methods, it has the
fewest number of wrongly assigned labels (row W). 
An accuracy of $92\%$, resp.~$99\%$ is reached at $\le 5$ intervening WordNet levels.

\section{Conclusions}
Among the other three methods, Markov Clustering (MCL) of the link graph
outperforms both MCL on the original graph and curvature
clustering. The number of wrongly assigned labels is about half of those
for {\em curv} and {\em orig} and the values in the 12  rows
are consistently higher with an accuracy of over $85\%$ at $\le
6$ WordNet levels.
The link graph clustering therefore
produces more accurate labels.
In the top frequency category, MCL on the
original graph has a slightly higher percentage of correctly assigned
class labels for small numbers of intervening WordNet levels, but is
soon overtaken by the link graph clustering.
The lower values for the curvature clustering can be partly explained by 
its low coverage. $854$ of the $1,200$ test words were not covered by the curvature clustering 
and had to be traced to clusters using depth-first search in 1 to 46 steps (with $60\%$ ($80\%$) of the test words being at most $3$ ($6$) links apart from a cluster). 

Judging by the classes in Table \ref{example_clusters}, we expect
curvature clustering to do especially well in recognizing the meanings
of words unknown to WordNet. 
%%%\marginpar{give some examples of
%%%non-WordNet words which got nicely mapped into the correct cluster}
%%%Link clustering, on the other hand, is especially well-suited in
%%%recognizing new meanings of ``old'' words.
%%%
\begin{table}[htpb]
\begin{center}
\caption[table_of_results_nn1]{\label{table_of_results_nn1} Evaluation results for common nouns}
\begin{tiny}
\begin{supertabular}{|l||p{0.7cm}|p{0.7cm}|p{0.7cm}|p{0.7cm}|}
\hline
& \multicolumn{4}{|c|}{nn1\_top} \\
\hline
Not & 2(0.01)  & 2(0.01)  & 2(0.01)  & 2(0.01) \\
Wrong & 74(0.37)  & 32(0.16)  & 69(0.34)  & 14(0.07) \\
1 & 21(0.17)  & 23(0.14)  & 8(0.06)  & 38(0.21) \\
2 & 40(0.32)  & 47(0.28)  & 26(0.20)  & 62(0.34) \\
3 & 66(0.53)  & 79(0.48)  & 50(0.39)  & 84(0.46) \\
4 & 90(0.73)  & 106(0.64)  & 70(0.54)  & 99(0.54) \\
5 & 99(0.80)  & 128(0.77)  & 89(0.69)  & 183(0.99) \\
6 & 108(0.87)  & 143(0.86)  & 103(0.80)  & 183(0.99) \\
7 & 114(0.92)  & 153(0.92)  & 110(0.85)  & 183(0.99) \\
8 & 118(0.95)  & 161(0.97)  & 120(0.93)  & 184(1.00) \\
9 & 121(0.98)  & 164(0.99)  & 124(0.96)  & 184(1.00) \\
10 & 121(0.98)  & 164(0.99)  & 127(0.98)  & 184(1.00) \\
11 & 122(0.98)  & 166(1.00)  & 129(1.00)  & 184(1.00) \\
12 & 123(0.99)  & 166(1.00)  & 129(1.00)  & 184(1.00) \\
\hline
& \multicolumn{4}{|c|}{nn1\_mid} \\
\hline
Not & 0(0.00)  & 0(0.00)  & 0(0.00)  & 0(0.00) \\
Wrong & 65(0.33)  & 38(0.19)  & 77(0.39)  & 25(0.12) \\
1 & 19(0.14)  & 17(0.11)  & 9(0.07)  & 33(0.19) \\
2 & 36(0.27)  & 45(0.28)  & 22(0.18)  & 62(0.35) \\
3 & 63(0.47)  & 74(0.46)  & 43(0.35)  & 80(0.46) \\
4 & 86(0.64)  & 105(0.65)  & 60(0.49)  & 90(0.51) \\
5 & 100(0.75)  & 127(0.79)  & 76(0.62)  & 174(0.99) \\
6 & 111(0.83)  & 139(0.86)  & 93(0.76)  & 174(0.99) \\
7 & 124(0.93)  & 150(0.93)  & 113(0.93)  & 174(0.99) \\
8 & 128(0.96)  & 156(0.97)  & 118(0.97)  & 174(0.99) \\
9 & 130(0.97)  & 158(0.98)  & 120(0.98)  & 174(0.99) \\
10 & 132(0.99)  & 159(0.99)  & 121(0.99)  & 174(0.99) \\
11 & 134(1.00)  & 161(1.00)  & 122(1.00)  & 174(0.99) \\
12 & 134(1.00)  & 161(1.00)  & 122(1.00)  & 174(0.99) \\
\hline
& \multicolumn{4}{|c|}{nn1\_low} \\
\hline
Not & 4(0.02)  & 4(0.02)  & 4(0.02)  & 4(0.02) \\
Wrong & 65(0.33)  & 47(0.23)  & 82(0.41)  & 38(0.19) \\
1 & 17(0.13)  & 13(0.09)  & 6(0.05)  & 20(0.13) \\
2 & 31(0.24)  & 33(0.22)  & 14(0.12)  & 32(0.20) \\
3 & 43(0.33)  & 64(0.43)  & 26(0.23)  & 50(0.32) \\
4 & 69(0.53)  & 87(0.58)  & 52(0.46)  & 62(0.39) \\
5 & 93(0.71)  & 112(0.75)  & 68(0.60)  & 145(0.92) \\
6 & 102(0.78)  & 127(0.85)  & 75(0.66)  & 152(0.96) \\
7 & 116(0.89)  & 132(0.89)  & 93(0.82)  & 156(0.99) \\
8 & 120(0.92)  & 143(0.96)  & 97(0.85)  & 157(0.99) \\
9 & 125(0.95)  & 147(0.99)  & 104(0.91)  & 158(1.00) \\
10 & 129(0.98)  & 148(0.99)  & 109(0.96)  & 158(1.00) \\
11 & 130(0.99)  & 148(0.99)  & 111(0.97)  & 158(1.00) \\
12 & 131(1.00)  & 149(1.00)  & 114(1.00)  & 158(1.00) \\
\hline
\end{supertabular}
\end{tiny}
\end{center}
\end{table}
We have shown that graphs can be learned directly from free text and
used for ambiguity recognition and lexical acquisition. We 
introduced two new combinatoric techniques, {\it graph curvature} and
{\it link clustering}, and evaluated their contribution as clustering
methods for lexical acquisition. Link clustering produces
particularly promising results when compared with information in the
WordNet noun hierarchy. These results demonstrate that our combinatoric
methods for analysing the geometry and topology of graphs improve
language learning.

\small
\bibliographystyle{acl}
\bibliography{sample}

\begin{thebibliography}{}

\bibitem[\protect\citename{Chartrand}1985]{chartrand-graph}
G.~Chartrand.
\newblock 1985.
\newblock {\em Introductory Graph Theory}.
\newblock Dover.

\bibitem[\protect\citename{Deerwester \bgroup et al.\egroup }1990]{Dee90}
S.~Deerwester, S.~Dumais, G.~Furnas, T.~Landauer, and R.~Harshman.
\newblock 1990.
\newblock Indexing by latent semantic analysis.
\newblock {\em Journal of the American Society for Information Science},
  41(6):391--407.

\bibitem[\protect\citename{Dorow and Widdows}2003]{Dor03}
B.~Dorow and D.~Widdows.
\newblock 2003.
\newblock Discovering corpus-specific word-senses.
\newblock In {\em Proceedings of EACL}, pages Conference Companion pp. 79--82,
  Budapest, Hungary, April.

\bibitem[\protect\citename{Eckmann and Moses}2002]{Eck02}
J.-P. Eckmann and E.~Moses.
\newblock 2002.
\newblock Curvature of co-links uncovers hidden thematic layers in the
  world-wide web.
\newblock In {\em Proceedings of the Natl. Acad. Sci. USA}, volume~99, pages
  5825--5829.

\bibitem[\protect\citename{Hearst and Sch\"utze}1993]{hearst-customizing}
M.~Hearst and H.~Sch\"utze.
\newblock 1993.
\newblock Customizing a lexicon to better suit a computational task.
\newblock In {\em ACL SIGLEX Workshop}, Columbus, Ohio.

\bibitem[\protect\citename{Hearst}1992]{hearst-hypernyms}
M.~Hearst.
\newblock 1992.
\newblock Automatic acquisition of hyponyms from large text corpora.
\newblock In {\em COLING}, Nantes, France.

\bibitem[\protect\citename{Pantel and Lin}2002]{Pan02}
P.~Pantel and D.~Lin.
\newblock 2002.
\newblock Discovering word senses from text.
\newblock In {\em Proceedings of ACM SIGKDD 2002}, Edmonton, Canada.

\bibitem[\protect\citename{Pereira \bgroup et al.\egroup }1993]{Per93}
F.~Pereira, N.~Tishby, and L.~Lee.
\newblock 1993.
\newblock Distributional clustering of english words.
\newblock In {\em Proceedings of ACL}, pages 183--190, Columbus, Ohio.

\bibitem[\protect\citename{Riloff and Shepherd}1997]{Ril97}
E.~Riloff and J.~Shepherd.
\newblock 1997.
\newblock A corpus-based approach for building semantic lexicons.
\newblock In {\em Proceedings of the Second Conference on Empirical Methods in
  NLP}, pages 117--124. ACL, Somerset, New Jersey.

\bibitem[\protect\citename{Roark and Charniak}1998]{Roa98}
B.~Roark and E.~Charniak.
\newblock 1998.
\newblock Noun-phrase co-occurence statistics for semi-automatic semantic
  lexicon construction.
\newblock In {\em COLING-ACL}, pages 1110--1116.

\bibitem[\protect\citename{Sch{\"u}tze}1998]{Sch98}
H.~Sch{\"u}tze.
\newblock 1998.
\newblock Automatic word sense discrimination.
\newblock {\em Computational Linguistics}, 24(1):97--124.

\bibitem[\protect\citename{Sigman and Cecchi}2002]{Sig02}
M.~Sigman and G.~Cecchi.
\newblock 2002.
\newblock The global organization of the wordnet lexicon.
\newblock In {\em Proceedings of the Natl. Acad. Sci. USA}, volume~99, pages
  1742--1747, February.

\bibitem[\protect\citename{Sproat and van Santen}1998]{Spr98}
R.~Sproat and J.~van Santen.
\newblock 1998.
\newblock Automatic ambiguity detection.
\newblock In {\em Proceedings of ICSLP 98}, Sydney, Australia.

\bibitem[\protect\citename{van Dongen}2000]{Don00}
S.~van Dongen.
\newblock 2000.
\newblock {\em Graph Clustering by Flow Simulation}.
\newblock {Ph.D.} thesis, University of Utrecht, May.

\bibitem[\protect\citename{Widdows and Dorow}2002]{Wid02}
D.~Widdows and B.~Dorow.
\newblock 2002.
\newblock A graph model for unsupervised lexical acquisition.
\newblock In {\em Proceedings of Coling}, pages 1093--1099, Taipei, Taiwan,
  August.

\bibitem[\protect\citename{Widdows}2003]{Wid03}
D.~Widdows.
\newblock 2003.
\newblock Unsupervised methods for developing taxonomies by combining syntactic
  and statistical information.
\newblock HLT-NAACL, Edmonton, Canada.

\end{thebibliography}

\end{document}